  \documentclass[prl,twocolumn,showpacs,superscriptaddress]{revtex4}

 \usepackage{amsmath} 
  \usepackage{graphicx,verbatim,mathenv}
 
  \def\gl         {\lambda}
  \def\w         {\omega}
  \def\gr         {\rho}

  \def\dk         {\frac{d\,{\bf k}}{\(2 \pi\)^3}}
  \def\dk1        {\frac{d\,{\bf k}_1}{\(2 \pi\)^3}}

  \def\beq        {\begin{equation}}
  \def\eeq        {\end{equation}}
  \def\bv         {{\mid}}

  \renewcommand{\(}{\left(}
  \renewcommand{\)}{\right)}

\begin{document}
  \title{Transforming nonlocality into frequency dependence: a shortcut to spectroscopy}
  \author{Matteo Gatti}
  \affiliation{Laboratoire des Solides Irradi\'es, \'Ecole Polytechnique, CNRS-CEA/DSM,  F-91128 Palaiseau, France}
  \affiliation{European Theoretical Spectroscopy Facility (ETSF)}
  \author{Valerio Olevano}
  \affiliation{Institut N\'eel, CNRS, F-38042 Grenoble, France}
  \affiliation{European Theoretical Spectroscopy Facility (ETSF)}
  \author{Lucia Reining}
  \affiliation{Laboratoire des Solides Irradi\'es, \'Ecole Polytechnique, CNRS-CEA/DSM,  F-91128 Palaiseau, France}
  \affiliation{European Theoretical Spectroscopy Facility (ETSF)}
  \author{Ilya V. Tokatly}
  \affiliation{Lehrstuhl f\"ur Theoretische Festk\"orperphysik, Universit\"at Erlangen-N\"urnberg, Staudtstrasse 7/B2, D-91054 Erlangen, Germany}
  \date{\today}
  \begin{abstract}
  
Measurable spectra are theoretically very often derived from complicated many-body Green's functions. In this way, one calculates much more information than actually needed. Here we
present an in principle exact approach to construct effective
potentials and kernels for the direct calculation of electronic spectra. In
particular, the potential that yields the spectral function needed to
describe photoemission turns out to be dynamical but {\it local} and
{\it real}. As example we illustrate this ``photoemission potential'' for sodium and aluminium, 
modelled as homogeneous electron gas, and discuss in particular its frequency dependence stemming from 
 the nonlocality of the
corresponding self-energy.  We also show that our approach leads to a
very short derivation of a kernel that is known to well describe
 absorption and energy-loss spectra of a wide range of materials.

  \end{abstract}
 
 

  \maketitle

The calculation of electronic excitations and spectra is one of the major challenges of today's condensed matter physics. 
In fact, while Density-Functional Theory (DFT), especially in simple and very efficient approximations like the Local-Density Approximation (LDA) \cite{KS},  has led to a breakthrough concerning the simulation of ground-state properties, the determination of electronic excited states is still quite cumbersome. Static ground-state DFT is not the adequate approach to spectroscopy. On the other hand, the
 full solution of the many-body Schr\"odinger equation for systems
of more than some tens of atoms is today out of reach. Clearly, the description of spectroscopy calls for the definition of suitable fundamental quantities (beyond the static ground-state density DFT is built on), and the derivation of new approximations. For neutral excitations (as measured e.g. in absorption or electron energy-loss spectroscopies) one can in principle work with the {\it time-dependent} density, and then try to approximate this Time-Dependent DFT (TDDFT), for example in the adiabatic LDA (TDLDA). 
Also Many-Body Perturbation Theory
(MBPT) approaches \cite{fetter}, like the GW approximation \cite{hedin} for the
one-particle Green's function or the Bethe-Salpeter equation (BSE) for
the determination of neutral excitations, are today  standard
methods for first-principles calculations of electronic excitations
\cite{RMP-NOI}.  One of the main reasons for their success 
is their intuitive physical picture that allows one to find
working strategies and approximations. Therefore, Green's functions
approaches are often used even when one could {\it in principle}
resort to more simplified methods, in particular density-functional
based ones - for example, when TDLDA breaks down.  It is hence quite logical to try to link both
frameworks \cite{fabien}, and derive better approximations for the
simpler (e.g. density-functional) approaches from the working
approximations of the more complex (e.g. MBPT) ones. Two prominent
examples for such a procedure are the optimized effective potential
(OEP) method \cite{OEP} where a local and static KS potential is
obtained from an in principle nonlocal and frequency dependent
self-energy, and recent derivations in TDDFT of a
two-point linear response xc kernel from the four-point BSE
\cite{kernel}. These constitute, at least at first sight, distinct
methods that solve particular questions and leave other, very
important points unsolved. The best example for such an open problem
is probably the so-called ``band-gap problem'' \cite{godby} of the Kohn-Sham (KS) \cite{KS}
approach: one would like to have a simple potential that yields not
only the correct density, but also band gaps in agreement with
experiment. Although OEP in the exact-exchange
approximation (EXX) \cite{EXX} had yielded some promising results in
this direction, it is clear that a systematic solution has not yet
been found \cite{myrta}.

In the present work we propose a general framework for the definition
of reduced potentials or kernels, designed for obtaining certain
quantities that can otherwise be calculated from some $N$-particle
Green's function. We show that both the abovementioned OEP and the
TDDFT kernel from MBPT are particular cases of our approach. As a new
level, we explore the question of a local and frequency dependent real
potential that allows one to find the correct trace of the
one-particle spectral function and therefore access direct and inverse
photoemission (including the bandgap).  We present results of this new
approach for  simple metals and a model insulator, and discuss the relation
to dynamical mean-field theory (DMFT) \cite{kotliar2}. Of
particular interest is the conversion of nonlocality into frequency
dependence that one encounters when an effective potential or kernel
with a reduced number of spatial degrees of freedom is used to
describe excitations.

In this general scheme we now suppose that we want to calculate the
quantity $T$ that is a part of the information carried by the
$N$-particle Green's function $G$. We symbolically express this
relation as: $T=p\{G\}$.  In the following we will
specify the formula for the one-particle $G$, whereas the completely
analogous case of two-particle Green's functions will be briefly
discussed at the end.  As a well-known example we take the electronic
density $T=\gr$ for which the ``part'' to be taken is the diagonal of
the one-particle $G$: $\gr (r) = -iG (r,r,t,t^+)$.  We then introduce
another Green's function $G_T$ which has the part $p\{.\}$ in
common with $G$: $T = p\{G_T\}$. We also suppose that $G_T$ is associated
to an effective potential $V_T$ according to $G_T = (\w - H_0 -
V_T)^{-1}$, where $H_0$ is the Hartree Hamiltonian \footnote{Here we
implicitly assume that $G$ and $G_T$ yield the same density, 
but this hypothesis can be 
straightforwardly generalized.}.  The full Green's function $G$ and
the new $G_T$ are linked by a Dyson equation:
\begin{equation}
\label{eq:dyson}
G = G_T + G_T \, (\Sigma - V_T) \, G.
\end{equation}
We now take the part of interest $p\{.\}$ of this Dyson equation. This
yields the condition:
\begin{equation}
p\{G_T \, (\Sigma - V_T) \, G \}=0.
\label{eq:dyson-part}
\end{equation}
The aim is now to make an {\it ansatz} for $V_T$ with a simpler structure
than $\Sigma$, and for which (\ref{eq:dyson-part}) can be solved.
For the example where $T$ is the density $\gr$, a static and local
potential can do the job. 
In fact in this example $V_T$ is the exchange-correlation potential
$V_{xc}(r)$, $G_{T}$ is the Kohn-Sham Green function $G_{KS}$, and
(\ref{eq:dyson-part}) is a well-known result, derived by Sham and
Schl\"uter in \cite{SSE}.  It has subsequently been extended to
time-dependent external potentials \cite{vanLeeuwen}, and employed in
many different contexts (see e.g. Refs. \cite{godby}\cite{ilya}).  The
Sham-Schl\"uter equation (SSE) is still implicit since self-consistent
in $V_{xc}$ and $G_{KS}$. Most often it is linearized setting $G =
G_{KS}$ everywhere, including in the construction of $\Sigma$. This so
called linearized Sham-Schl\"uter equation, that can also be derived
using a variational principle \cite{casida}\cite{ulf}, is the central
equation of the OEP method \cite{OEP}.  In particular, if one uses for
$\Sigma$ the exchange-only approximation $\Sigma=\Sigma_x$, where
$\Sigma_x = iG_{KS}v$ ($v$ is the bare Coulomb interaction), one
obtains the so called exact-exchange approximation to $V_{xc}$
\cite{EXX}.  Obviously, by construction one obtains in this way a good
description of the density, whereas there is no reason for other
features of the Green's function, in particular the band gap, to be
correct.  However, in our much more general scheme one can easily go
beyond the case of solely the density $\gr(r)$. 
According to the choice of $p\{.\}$, the {\it ansatz} for $V_T$ has then
to be modified.
Below we
will first explore the problem of electron addition and removal.

Electronic structure, defined as electron addition and removal
energies, is measured by experiments like direct or inverse
photoemission.  To first approximation \cite{RMP-NOI} these
experiments measure the trace of the spectral function $A(r_1,r_2,\w)
= \frac{1}{\pi} \bv \textrm{Im} \, G(r_1,r_2,\w) \bv $. In other words,
for the interpretation of photoemission spectra one doesn't need the
knowledge of the whole Green's function $G$, but just the imaginary
part of its trace over spatial coordinates together with its full {\it
frequency dependence}.  We will hence look for a potential that is
simpler than the full self-energy but yields the correct trace of the
spectral function. It is reasonable to add the condition that also the
density should be correct, which means that the diagonal in real
space, and not only its integral, is fixed. Following the general
scheme, we introduce a new Green's function $G_{SF}=(\w -
H_{0}-V_{SF})^{-1}$ stemming from a potential $V_{SF}$ such that:
$\textrm{Im} G_{SF}(r_1,r_1,\w) = \textrm{Im} G(r_1,r_1,\w)$.  What
degrees of freedom are needed in $V_{SF}$? A natural assumption is
that $V_{SF}$ should be local in space, but frequency dependent
\cite{qplda1}\cite{qplda2}. It is also possible to choose $V_{SF}$ to
be {\it real}. With this {\it ansatz} Eq. (\ref{eq:dyson-part})
yields:
\begin{multline}
V_{SF} (r_1,\w) =  \int  dr_2dr_3dr_4 \, \xi^{-1}(r_1,r_4,\w) \,
\times 
\\ \textrm{Im} \{ G_{SF}(r_4,r_2,\w)\Sigma(r_2,r_3,\w)G(r_3,r_4,\w) \} 
\label{defVSF}
\end{multline}
where $\xi(r_1,r_2,\w) = \textrm{Im} \{
G_{SF}(r_1,r_2,\w)G(r_2,r_1,\w) \}$.  The solution exists if $\xi$ is
invertible. Whether this is always the case is a delicate question and
beyond the scope of the present work.  Equation (\ref{defVSF}) shows
that $V_{SF}$ should indeed be frequency dependent unless $\Sigma$ is
{\it static} and {\it local} (in that case the $\omega$-dependent
terms cancel trivially). Thus, in general a local static (KS)
potential will not be able to reproduce the spectral function, whereas
$V_{SF}(r,\omega)$ {\it is} the local potential that will yield the
correct bandgap {\it and} the correct density $\gr(r)$ of the system.

At this point it is interesting to compare our construction
with the approach of the Spectral Density-Functional Theory (SDFT)
\cite{kotliar1}, where the key variable is the short-range part of the
Green's function: $G_{loc} (r,r',\w) = G(r,r',\w) \,
\Theta(\Omega_C)$, where $\Theta(\Omega_C)$ is $1$ when $r$ is in the
unit cell and $r'$ inside the volume $\Omega_C$, and is $0$ otherwise
(see Fig. 1 of Ref. \cite{kotliar1}).  A new Green's function
$G_{SDFT}=(\w - H_{0}-V_{SDFT})^{-1}$ can be introduced such that
$G_{SDFT} = G_{loc}$ where $G_{loc}$ is different from 0. Using this
property of $G_{SDFT}$, we find that the (in general complex)
potential $V_{SDFT}$, defined in the volume $\Omega_C$, is:
\begin{widetext}
\beq
V_{SDFT}(r_5,r_6,\w) = \int_{\Omega_C} \, dr_1 dr_2 \int dr_3 dr_4  \, \tilde{G}^{-1}_{SDFT}(r_5,r_1,\w) G_{SDFT}(r_1,r_3,\w)
\Sigma(r_3,r_4,\w) G(r_4,r_2,\w)  \tilde{G}^{-1}(r_2,r_6,\w)
\eeq
\end{widetext}
where $\tilde{G}^{-1}$, if it exists, is the local inverse of $G$ in
$\Omega_C$ (while $G^{-1}$ would be the full inverse, defined in the
whole space).  In principle SDFT is a formally exact theory. The most
common approximation to SDFT is the dynamical mean-field theory
\cite{kotliar2}. In this perspective \cite{kotliar1} DMFT can be
viewed as consisting in the assumption of taking for the interaction
energy functional of SDFT $\Phi_{SDFT}[G_{loc}]$ the form of
functional $\Phi_{MB}[G]$ for the full Green's function
$G$.  
This means that the shorter the range of the
self-energy is, the better this approximation becomes. In particular,
in the limit case that $\Sigma$ is completely localized in $\Omega_C$,
then $V_{SDFT}$ and $G_{SDFT}$ coincide respectively with $\Sigma$ and
$G$.  The interesting situation is of course when this is not true. In
fact, our $V_{SF}$ of Eq. (\ref{defVSF}) corresponds to the case where
$\Omega_C \rightarrow 0 $ so that this condition is certainly not
fulfilled. Then, as we will illustrate below, the nonlocality of
$\Sigma$ will strongly influence $V_{SF}$ and in particular lead to a
frequency dependence which is {\it not} the frequency dependence of
$\Sigma$ itself. This will to a certain extent also be true for any
$\Omega_C $ of finite range, so that the following discussions may
also give useful insight for research in the field of DMFT.

\begin{figure}
\begin{center}
\includegraphics[width=\columnwidth]{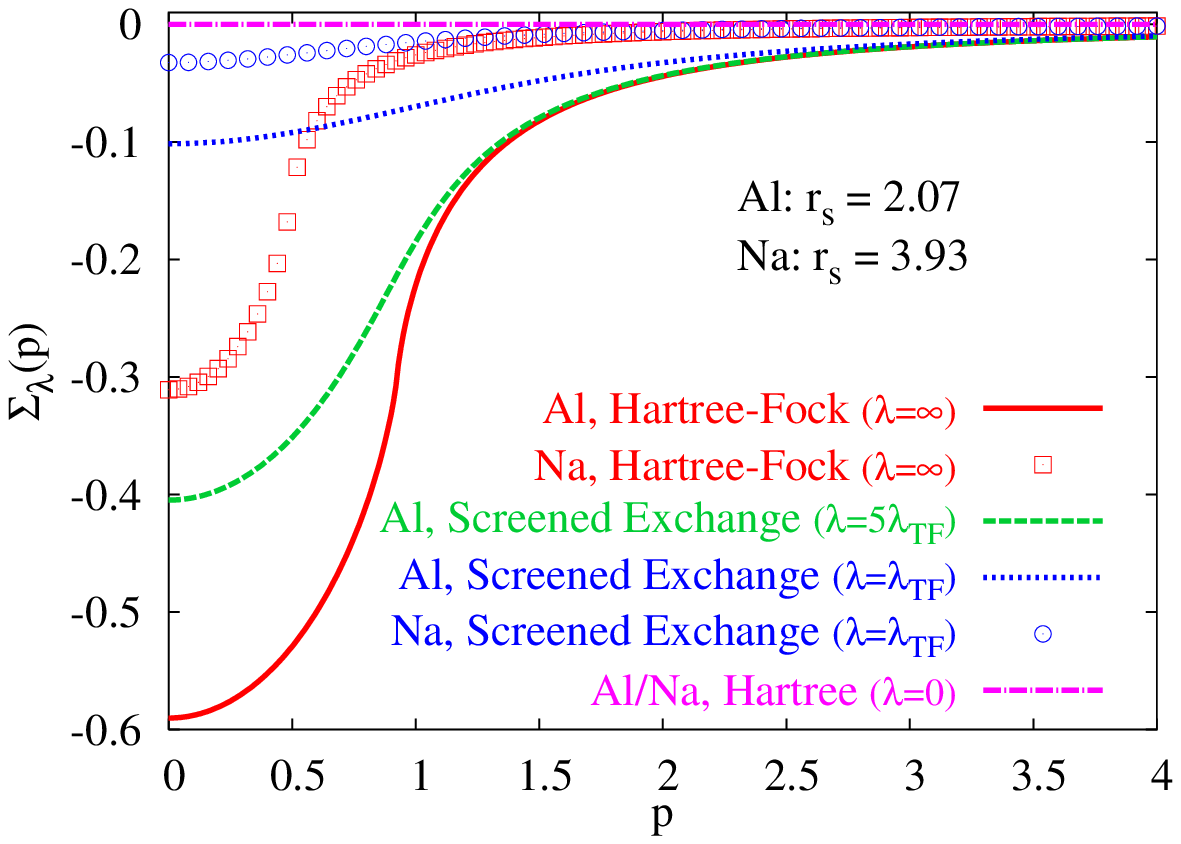}\\
\includegraphics[width=\columnwidth]{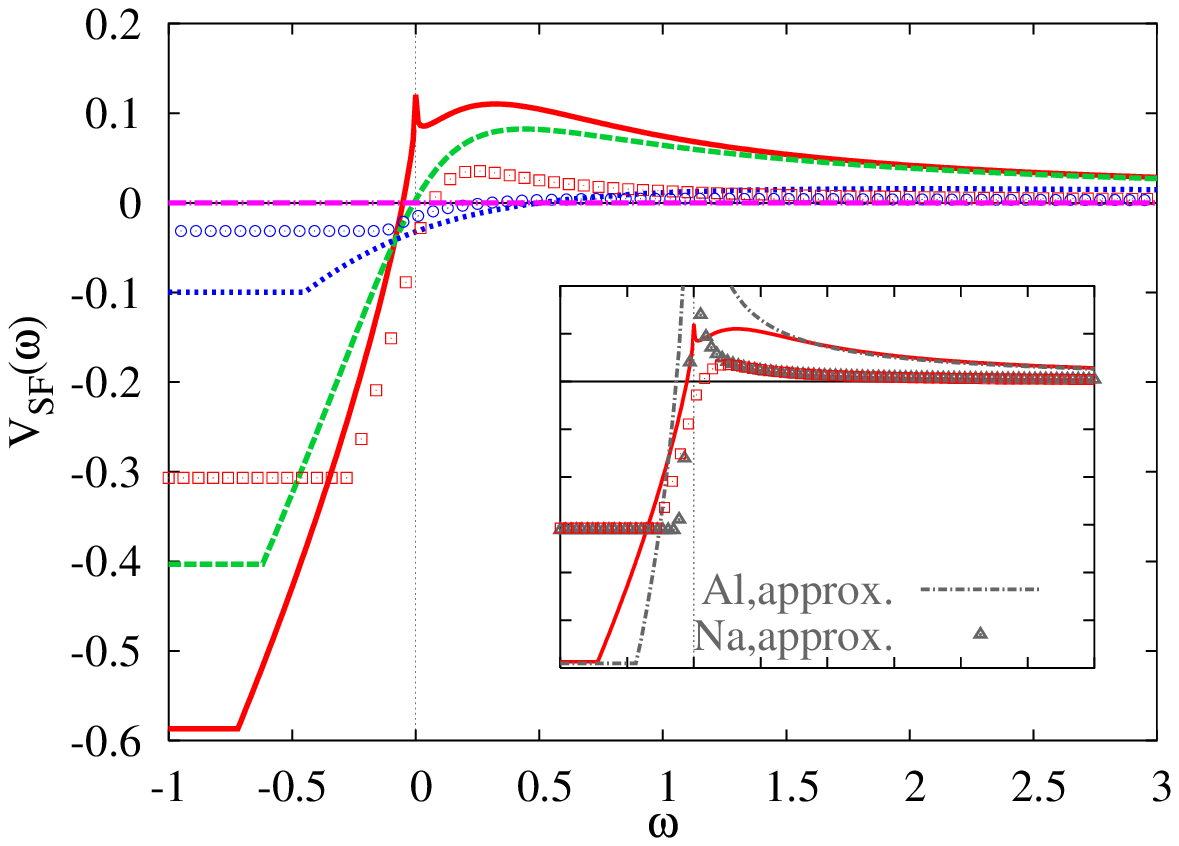}
\end{center}
\caption{(Color online) Transformation of  nonlocal statically screened exchange
  self-energy $\Sigma_{\gl}(p)$ (upper panel) to the frequency dependent local
  $V_{SF}(\w)$ (bottom panel), for different screening lengths $\gl$
  in HEG ($\lambda_{TF}=(4p_F/\pi)^{-1/2}$ is the Thomas-Fermi length), and densities 
  corresponding to aluminium (lines) and sodium (open symbols). 
  The key is common to both panels. 
  The $\w$-dependence of $V_{SF}$  is stronger for more nonlocal self-energies. Below
  band-edges, where $\textrm{Im}G=0$, $V_{SF}(\w)$ has been defined
  continuous and equal to a constant $>\w+\mu$. In the inset:
  comparison between the exact and approximated solutions
  (\protect\ref{Vexact}) and  (\protect\ref{vsf_ghgh}) in the
  Hartree-Fock case.} 
\label{fig2}
\end{figure}

To illustrate the frequency dependence of $V_{SF}$ we
consider the case of homogeneous systems, where all local
quantities (like $V_{xc}$ and $V_{SF}$) are constant in space.  In
particular, the xc potential is $V_{xc} =
\Sigma(p=p_F,\w=0)$, while, since $V_{SF}=V_{SF}(\w)$, one can
directly write (here and throughout the paper we adopt atomic units):
\begin{multline}
\bv\textrm{Im}G_{SF}(r,r,\w)\bv =
2\int\frac{d_3p}{8\pi^3}\pi\delta(\w+\mu-p^2/2-V_{SF}(\w)) \\
=\frac{\sqrt{2}}{\pi}\theta(\w+\mu-V_{SF}(\w))\sqrt{\w+\mu-V_{SF}(\w)}. 
\label{eq:g-heg-sf}
\end{multline}
Requiring that this is equal to $\bv \textrm{Im} G(r,r,\w) \bv $, one
finds a unique local potential $V_{SF}$ (for $V_{SF}(\w)<\w+\mu$):
\begin{equation}
V_{SF}(\w) = \w+\mu  -  \Big( \frac{\pi}{\sqrt{2}}  \, \bv \textrm{Im}
G(r,r,\w) \bv \Big)^2. 
\label{Vexact}
\end{equation}
It is first of all interesting to consider the case of simple metals, that can be modelled by a
 homogeneous electron gas. Here, we will assume 
 a static but nonlocal self-energy \footnote{In order to make a comparison 
with experimental results we should consider self-energies that include dynamical correlations. 
Such an additional frequency dependence in $\Sigma$  would trivially add to the frequency dependence of $V_{SF}$.}:
$\Sigma_{\gl}(r-r')=iG(r-r',t-t^+)v_{\gl}(r-r')$, that is a
screened-exchange like form where the screened Coulomb potential is
$v_{\gl}(r-r') = v(r-r') e^{-\bv r-r' \bv/\gl}.$ For a larger
screening length $\gl$, $\Sigma$ is more effectively nonlocal, since
less screened.  $\gl$ tunes the effective range of the interaction:
from $\gl=0$ ($\Sigma=0$, Hartree approximation), to $\gl = \infty$
(unscreened Hartree-Fock).
Using the relation
(\ref{Vexact}) we calculate $V_{SF}(\w)$ for different
values of $\gl$, i.~e. for different nonlocality ranges of $\Sigma$
(see Fig. \ref{fig2}).
We find that the spatial nonlocality of the static self-energies is
completely transformed into the frequency dependence of $V_{SF}$.  
This is an essential property of $V_{SF}$, which radically
distinguishes $V_{SF}$ from the static $V_{xc}$.  
In particular, looking at Fig. \ref{fig2}, one observes that {\it the more
nonlocal $\Sigma_{\gl}$ is, the more dynamical $V_{SF}$ becomes}.   
This is confirmed by comparing two materials: aluminium and sodium.

The HEG is a prototype of metallic systems. Let us consider the
simplest model insulator introduced by Callaway
\cite{Callaway1959}. The Callaway's model is obtained from HEG by
inserting a gap $\Delta$ between the occupied ``valence'' states
$\phi_{v}(r)$ ($p<p_{F}$) and the ``conduction'' states $\phi_{c}(r)$
($p>p_{F}$). Within the Green's functions formalism this corresponds
to a nonlocal static ``scissor'' self-energy
\beq 
\Sigma(r,r') = 
\Delta \sum_c \phi_c(r)\phi_c^*(r'), 
\label{scissor} 
\eeq 
which rigidly shifts all conduction states, and produces a gap between
the valence, $E_{v}=\varepsilon_{F}=\mu$, and the conduction,
$E_{c}=\varepsilon_{F}+\Delta$, band edges. Due to the homogeneity of the
model, Eq. (\ref{Vexact}) is still applicable.
From Eq. (\ref{Vexact}), one gets that 
$V_{SF}(\w) = 0$ for $\w < E_v-\mu = 0$ and $V_{SF}(\w) =\Delta$ for
$\w > E_c-\mu = \Delta $.  Between $\w =0$ and $\w= \Delta $, where
$\textrm{Im} G(\w)=0$, there are many different choices for
$V_{SF}(\w)$: from (\ref{eq:g-heg-sf}) it is enough that
$V_{SF}(\w)\ge\w+\mu$ to get $\textrm{Im} G_{SF}(\w)=0$.
In any case a static constant potential, like $V_{xc}$, cannot produce
the correct bandgap, because at $E_v-\mu$ and $E_c-\mu$ it should
assume different values. 
The only effect of $V_{xc}$ with respect to $H_0$ would be just a
rigid shift of the whole bandstructure and no gap could be opened.  In
fact, $\Delta$ in (\ref{scissor}) corresponds to the derivative
discontinuity of $V_{xc}$ when an electron is added to the system:
$\Delta = V^{(N+1)}_{xc}(r) - V_{xc}^{(N)}(r)$ \cite{godby}.  Our
simple example demonstrates that this discontinuity, which constitutes the
difference between the true quasiparticle gap and the Kohn-Sham one,
is accounted for by the potential $V_{SF}$ rightly through its
frequency dependence.

Knowing the exact solutions, one can also verify how common
approximations to SSE perform with these strongly frequency dependent
functions, in particular when we linearize the SSE by setting
$G=G_{SF}$ everywhere (in DFT this would correspond to the OEP
approach).  The linearized SSE is still self-consistent in $V_{SF}$
and $G_{SF}$.  Yet we can approximate it at the first order setting
$G$ and $G_{SF}$ equal to the Hartree $G_H$. In the Hartree-Fock case
in HEG, we get, for $\w> -\varepsilon_F$:
\beq
V_{SF}\Big(\tilde{\w}=\frac{\w+\varepsilon_F}{\varepsilon_F}\Big) =
-\frac{2p_F}{\pi} \Big( 1 + \frac{\sqrt{\tilde{\w}}}{2} \ln
\Big|\frac{\sqrt{\tilde{\w}}-1}{\sqrt{\tilde{\w}}+1}\Big| \Big)
\label{vsf_ghgh}
\eeq
which shows a logarithmic divergence at the Fermi energy $\w=0$ and
hence is very different from the exact solution (see the inset of
Fig. \ref{fig2}).  The approximated $V_{SF}$ in HEG is distant from
the exact one also when the linearized SSE is solved
self-consistently.  The agreement should improve when screening is
taken into account, since then $V_{SF}$ is less frequency dependent
and $G_{SF}$ is much closer to $G$ and $G_H$. 
Already for sodium the agreement is better than for aluminium.
 In any case, these
results demonstrate that the linearization of the generalized SSE is a delicate
procedure.  In the following we will however show that it works very well
in certain cases, for example for the kernel $f_{xc}$  of TDDFT.

Up to now we have considered Dyson equations only for the one-particle
Green's function. The generalization of SSE can however be applied
to any Dyson-like equation, for example involving two-particle Green's
functions, like the four-point reducible polarizability
${}^4\chi_{MB}$.  In this way we can derive an exact equation for the
kernel $f_{xc}=f_{xc}^{(1)}+f_{xc}^{(2)}$ of TDDFT. In particular, following Ref. \cite{fabien},
we concentrate on the term $f_{xc}^{(2)}$ that describes
electron-hole interactions. This means that in the TDDFT
polarizability ${}^4\chi_{TD}$ the ``gap opening'' contribution due to
the term $f_{xc}^{(1)}$ is already included by using  $\chi_0^{QP}=-iGG$ 
instead of the KS independent particle polarizability. $f_{xc}^{(2)}$ is the
two-point kernel that gives the correct {\it two-point} polarizability
\footnote{One has to be careful about the link between time-ordered
and retarded quantities.}: ${}^4\chi_{TD}(1122) = {}^4\chi_{MB}(1122)
\equiv \chi(12)$. Applying this condition 
to the Dyson equation ${}^4\chi_{MB} = {}^4\chi_{TD}[F_{MB}-f_{xc}^{(2)}]\,{}^4\chi_{MB}$
yields the generalized SSE, and hence the exact expression of $f_{xc}^{(2)}$:
\begin{multline}
f_{xc}^{(2)}(34) =  \chi^{-1}(31) {}^4\chi_{TD}(1156)F_{MB}(5678) \times  \\
\times {}^4\chi_{MB}(7822)\chi^{-1}(24), 
\end{multline}
where $F_{MB}=i\delta\Sigma/\delta G$ is the exchange-correlation part
of the kernel of the Bethe-Salpeter equation, which, in the framework
of {\it ab initio} calculations \cite{RMP-NOI}, is most frequently
approximated by: $F_{MB}(1234)=-W^{st}(12)\delta(13)\delta(24)$ where
$W^{st}(12)=W(r_1,r_2,\w=0)\delta(t_2-t_1^+) $ is the statically
screened Coulomb interaction.  An approximated expression for
$f_{xc}^{(2)}$, obtained from the linearized generalized SSE, where we
set $\chi(12)= \chi_0^{QP}(12)$ and
$\chi^3(1;23)={}^4\chi_{MB}(1123)={}^4\chi_{TD}(1123)={}^4\chi_0^{QP}(1123)=-iG(12)G(31)$,
is then:
\beq
f_{xc}^{(2)}(\w) = (\chi_0^{QP})^{-1}(\w)\chi^3(\w)W(0) 
{}^3\chi (\w)(\chi_0^{QP})^{-1}(\w). 
\label{NQkernel}
\eeq
In this equation only the frequency dependence of the various terms
has been put into evidence, since it is mostly interesting to note
that, although $W$ and hence $F_{MB}$ are static, $f_{xc}$ is frequency
dependent unless $W$ is short-ranged in which case the frequency
dependence of the other components cancels.  This recalls the
analogous transformation of static self-energies into frequency
dependent potentials discussed above. Equation (\ref{NQkernel})
represents a new derivation of a two-point linear response
exchange-correlation kernel that has previously been obtained in
several other ways \cite{kernel}. It has been shown to yield
absorption or energy-loss spectra of a wide range of materials in very
good agreement with experiment \cite{kernel}. This means also that
the linearization of the generalized SSE in this case turns out to be a very good
approximation.  The present derivation is particularly quick and
straightforward, showing one of the advantages of the generalized
Sham-Schl\"uter equation formulation.

In conclusion, in this paper we propose a shortcut for the calculation of electronic spectra, based on a
generalization of the
Sham-Schl\"uter equation \cite{SSE}. In particular, we have introduced a {\it local} and {\it real} potential for photoemission, with a frequency dependence stemming both from the frequency dependence and from the {\it non-locality} of the underlying self-energy. 
We have illustrated some features at the example of sodium and aluminium. We have also applied the approach to the derivation of an exchange-correlation kernel for the calculation of absorption spectra.   This work opens the way to explore new approximations for
simplified potentials and kernels that can be employed to calculate a wide range of electronic spectra.

We are grateful for discussions with R. Del Sole, G. Onida,
F. Sottile, and F. Bruneval, 
and support from the EU's 6th
Framework Programme through the NANOQUANTA Network of Excellence
(NMP4-CT-2004-500198) and from ANR (project NT0S-3$\_$43900). 

  \bibliographystyle{apsrev}
 
  
  %
  %

  \end{document}